\documentclass[aps,prl,preprintnumbers,showpacs,nofootinbib,floatfix]{revtex4}
\usepackage{graphicx}% Include figure files
\usepackage{comment}
\usepackage{dcolumn}% Align table columns on decimal point
\usepackage{bm}% bold math
\begin{document}
\newcommand{\newc}{\newcommand}
\newc{\ra}{\rightarrow}
\newc{\lra}{\leftrightarrow}
\newc{\beq}{\begin{equation}}
\newc{\eeq}{\end{equation}}
\newc{\barr}{\begin{eqnarray}}
\newc{\earr}{\end{eqnarray}}
%%%%%%%%%%%%%%%%%%%%%%%%%%%%%%%%%%%%%%%%%%%
\newcommand{\Od}{{\cal O}}
\newcommand{\lsim}   {\mathrel{\mathop{\kern 0pt \rlap
  {\raise.2ex\hbox{$<$}}}
  \lower.9ex\hbox{\kern-.190em $\sim$}}}
\newcommand{\gsim}   {\mathrel{\mathop{\kern 0pt \rlap
  {\raise.2ex\hbox{$>$}}}
  \lower.9ex\hbox{\kern-.190em $\sim$}}}
%\preprint{APS/123-QED}

\title {COHERENT NEUTRAL CURRENT  NEUTRINO-NUCLEUS SCATTERING AT A SPALLATION SOURCE; A VALUABLE EXPERIMENTAL PROBE}
%\title {EXPLOITING  TPC DETECTORS AND THE COHERENT  NEUTRAL CURRENT INTERACTION
%FOR DETECTING  SUPERNOVA NEUTRINOS}
%
%
%\toctitle{ Neutrinos \protect\newline in a Spherical Box}
% allows explicit linebreak for the table of content
%
%
%\titlerunning{NEUTRINOS IN A SPHERICAL BOX}
% allows abbreviation of title, if the full title is too long
% to fit in the running head
%
\author{J.D. Vergados$^{1,2}$\thanks{Vergados@uoi.gr}, F.T.  Avignone III$^{3}$ and I. Giomataris$^{4}$  }
%
%\authorrunning{Giomataris and Vergados}
% if there are more than two authors,
% please abbreviate author list for running head\author{J. D. Vergados$^{(1),(2)}$\thanks{Vergados@uoi.gr} and  H. Ejiri$^{(3),(4),(5)}$\thanks{ejiri@rcnp.osaka-u.ac.jp}}

\affiliation{$^{(1)}${\it University of Ioannina, Ioannina, GR 45110, Greece .}}
\affiliation{$^{(2)}${\it Theory Division, CERN, CH,}}
\affiliation{$^{(3)}${\it University of South Carolina, Columbia, SC 29208, USA, }}
\affiliation{$^{(4)}${\it CEA, Saclay, DAPNIA, Gif-sur-Yvette, Cedex,France,}}
%
%\address{
%1 University of S, Carolina, Columbia, S.C., USA}
%\address{
%2 CEA, Saclay, DAPNIA, Gif-sur-Yvette, Cedex,France}
%\address{
%3 University of Ioannina, Ioannina, GR 45110, Greece
%\\E-mail:Vergados@cc.uoi.gr}
%\maketitle              % typesets the title of the contribution
\begin{abstract}
The coherent contribution of all neutrons in neutrino nucleus scattering due to the neutral current is examined considering the Spallation Neutron Source (SNS) as a source of neutrinos. SNS is a prolific pulsed source of
electron and muon neutrinos as well as muon antineutrinos.
\end{abstract}
%\begin{keyword}
%\end{keyword}
\date{\today}
%%%%%%%%%%%%%%%%%%%%%%%%%%%%%%%%%%%%%%%%%%%%%%%%%%%%%%%%%%%%%%%%%%%%%
\maketitle
\section{Introduction}
The question of detecting and exploiting neutrinos from both terrestrial and extra terrestrial sources
has become central to physics and astrophysics. In recent years, in particular,
 exploiting the neutral current interaction for detecting
neutrino induced nuclear recoils has become fashionable. Indeed
due to the coherence of all neutrons in the nucleus one gets very
large cross sections, which are proportional to the square of the
neutron number (the coherent neutral current contribution due to
protons is negligible). To exploit this feature, however,
 one should design large detectors with very low  energy thresholds. Two possibilities have been considered:
 \begin{itemize}
 \item Spherical TPC detectors filled with a noble gas for supernova neutrino detection.
 Such detectors
 can give us information about the neutrino source, since the neutral current interaction is not sensitive
 to detailed neutrino properties like neutrino masses, mixing etc, which appear in neutrino oscillations.
 It has also been pointed out that such detectors are robust, cheap and easy to maintain for years. So they
 can surely detect supernovae at distances of  tens of kiloparsecs \cite{VERGIOM06}.
 \item The standard direct dark matter detectors aim at detecting nuclear recoils due to WIMP's
 (weakly interacting massive particles). Since the expected event rates are extremely low, one may
have to worry about the rare background recoils due to neutrino sources,  like the solar $^8$B  neutrinos  \cite{VerEji08}, \cite{VEG08}.
 \end{itemize}
 It is thus important to study the behavior of such detectors using well known terrestrial
 neutrino sources. One such source  is  Spallation Neutron Source (SNS) at the Oak Ridge National Laboratory, Oak Ridge, Tennessee, which  is the most   powerful source of pulsed intermediate-energy neutrinos \cite{AVIGNONE},\cite{AVIGNONE1} .

\begin{figure}[!ht]
 \begin{center}
 \rotatebox{90}{\hspace{1.0cm} {$f_{\nu}(E_{\nu})\rightarrow $}}
\includegraphics[scale=0.8]{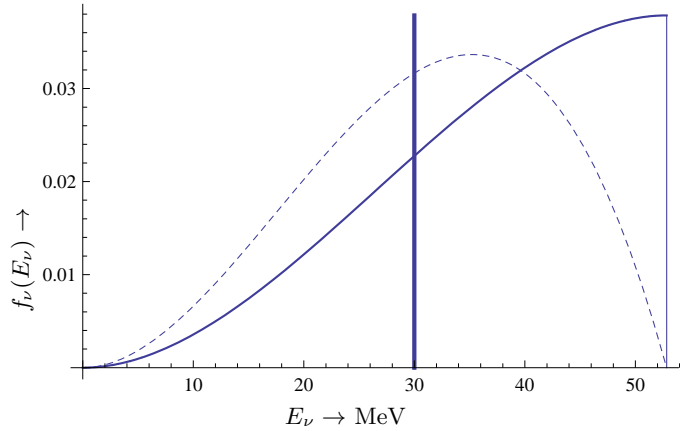}
\hspace{8.0cm}$E_{\nu} \rightarrow$ MeV
 \caption{The  neutrino spectrum from stopped pions. The normalized dashed  and solid curves correspond
to $\nu_e$ and $\tilde{\nu}_{\mu}$ respectively.  Also shown is the discreet $\nu_{\mu}$ spectrum (thick vertical line).}
 \label{fig:nuspec}
 \end{center}
  \end{figure}

 The primary parameters of the SNS are as follows:\\
Proton beam power on target:        1.4 MW\\
Proton beam kinetic energy:          1.0 GeV.\\
Average beam current on target:    1.4 mA\\
Protons per pulse on target:          $  1.5\times 10^{14}$\\
Pulse repetition rate:                        60 Hz\\
$\pi^+$   per proton:                               0.068\\
 $\pi^+$  per second                                  $6.14\times 10^{14}$\\
$R (\nu_e)=R(\bar{\nu}_{\mu})=                                6.14\times 10^{14}$\\
%\end{document}
The total neutrino flux at a distance $r$ is:
\beq
\Phi(\nu_e,r)=\Phi(\bar{\nu}_{\mu},r)=\Phi(\nu_{\mu},r)=\frac{6.14x10^{14} \mbox{s}^{-1}}{4 \pi r^2}
\eeq
At a typical distance of 50 m we find:
\beq
\Phi(\nu_e,r)=\Phi(\bar{\nu}_{\mu},r)=\Phi(\nu_{\mu},r)=1.95\times 10^6 \mbox{cm}^{-2}\mbox{s}^{-1}
\eeq
The normalized neutrino spectra can be described very well by the shapes exhibited in
Fig. \ref{fig:nuspec}
  \section{ Elastic Neutrino nucleon Scattering}
% For low energy neutrinos the historic process neutrino-electron scattering \cite{HOOFT} \cite{REINES}
% is very useful.
%The differential cross section \cite{VogEng} takes the form
%('t Hooft and Vogel $\&$ Engel)
%\begin{equation}
%\frac{d\sigma}{dT}=\left(\frac{d\sigma}{dT}\right)_{weak}+
%\left(\frac{d\sigma}{dT}\right)_{EM} \label{elas1a}
%\end{equation}
The cross section for elastic neutrino nucleon scattering has extensively been studied.
% \cite{ELNUNUC}.
It has been shown that at low energies it can be simplified and  be cast in the form:
\cite{BEACFARVOG},\cite{VogEng}:
 \begin{eqnarray}
 \left(\frac{d\sigma}{dT_N}\right)_{weak}&=&\frac{G^2_F m_N}{2 \pi}
 [(g_V+g_A)^2\\
\nonumber
&+& (g_V-g_A)^2 [1-\frac{T_N}{E_{\nu}}]^2
+ (g_A^2-g_V^2)\frac{m_NT_N}{E^2_{\nu}}]
 \label{elasw}
  \end{eqnarray}
  where $m_N$ is the nucleon mass, $T_N$ its energy and $g_V$, $g_A$ are the weak coupling constants. Neglecting their
  dependence on the momentum transfer to the nucleon they take the form:
  \beq
 g_V=-2\sin^2\theta_W+1/2\approx 0.04~,~g_A=\frac{1.27}{2} ~~,~~(\nu,p)
\label{gcoup1}
\eeq
\beq
g_V=-1/2~,~g_A=-\frac{1.27}{2}~~,~~(\nu,n)
 \label{gcoup2}
 \eeq
 In the above expressions for the axial current the renormalization
in going from the quark to the nucleon level was taken into account. For antineutrinos $g_A\rightarrow-g_A$.
To set the scale we write:
\beq \frac{G^2_F m_N}{2 \pi}=5.14\times 10^{-41}~\frac{\mbox{cm}^2}{\mbox{MeV}}
\label{weekval}
\eeq
% In the above expressions for the $\nu_{\mu},\nu_{\tau}$ only the
% neutral current has been included, while for $\nu_e$ both the
% neutral and the charged current contribute.
The nucleon energy depends on the  neutrino energy and the
scattering angle, the angle between the direction of the recoiling particle and that of the incident neutrino. In the laboratory frame  it is given by:
\beq
T_N= \frac{2~m_N (E_{\nu}\cos{\theta})^2}{(m_N+E_\nu)^2-(E_{\nu}
\cos{\theta})^2}~~,~~0\leq \theta\leq \pi/2
\label{eq:TN}
\eeq
%$$T=\frac{X^2}{2 m_e}~~,~~X=2E_{\nu} \frac{m_e(m_e+E_{\nu})\cos{\theta}}
%{(m_e+E_\nu)^2-(E_{\nu} \cos{\theta})^2}$$
(forward scattering). For sufficiently small neutrino energies, the last equation can be simplified as follows:
$$ T_N \approx \frac{ 2(E_\nu \cos{\theta})^2}{m_N}$$
The above formula can be generalized to any target and can be written in dimensionless form
as follows:
\beq
y=\frac{2\cos^2{\theta}}{(1+1/x_{\nu})^2-\cos^2{\theta}}~~,~~
y=\frac{T_{recoil}}{m_{recoil}},x_{\nu}=\frac{E_{\nu}}{m_{recoil}}
\label{recoilen}
\eeq
In the present calculation we will treat $x_{\nu}$ and $y$ as dynamical variables, in line with CDM recoils. One, of course,
equally well could have chosen $x_{\nu}$ and $\theta$ as relevant variables.

 The maximum  energy occurs when $\theta=0$, i.e.:
\beq
 y_{max}=\frac{2}{(1+1/x_{\nu})^2-1},
 \label{Eq:ymax}
 \eeq
in agreement with Eq. (2.5) of ref. \cite{BEACFARVOG}.
% This relationship is
%plotted in  Fig. \ref{fig:yofx}.
%I what follows, whenever appropriate, we are going to average our results with the neutrino spectrum
%shown in Fig. \ref{spectrum}.
% \begin{figure}[!ht]
% \begin{center}
%\rotatebox{90}{\hspace{0.0cm} {$\rightarrow \frac{T_{recoil}}{m_{recoil}}$}}
%\includegraphics[scale=0.3]{scale_ener1.eps}
%\hspace{1.0cm}$\rightarrow \frac{E_{\nu}}{m_{recoil}}$
% \rotatebox{90}{\hspace{0.0cm} {$\rightarrow \frac{T^{max}_{recoil}}{m_{recoil}}$}}
%\includegraphics[scale=0.8]{scale_ener1.eps}
%\hspace{6.0cm}$\rightarrow \frac{E_{\nu}}{m_{recoil}}$
% \rotatebox{90}{\hspace{0.0cm} {$\rightarrow \frac{T_{recoil}}{m_{recooil}}$}}
%\includegraphics[scale=0.3]{scale_ener2.eps}
%\hspace{2.0cm}$\rightarrow \frac{E_{\nu}}{m_{recoil}}$
% \caption{The maximum recoil energy  as a function of the neutrino energy (both in units of the
%recoiling mass). The scale is realistic for nuclear recoils of the type of experiments considered here.}
 %\label{fig:yofx}
 %\end{center}
 % \end{figure}
  One can invert Eq. \ref{recoilen} and get the  neutrino energy associated with a given recoil energy and
scattering angle. One finds
\beq
  x_{\nu}=\left[-1+\cos{\theta} \sqrt{1+\frac{2}{y}} \right]^{-1}~~,~~0\leq \theta\leq \pi/2
  \label{xofyxi}
  \eeq
  The minimum neutrino energy for a given recoil energy is given by:
    \beq
  x^{min}_{\nu}=\left[-1+\sqrt{1+\frac{2}{y}} \right]^{-1}=\frac{y}{2}(1+\sqrt{1+\frac{2}{y}})
  \label{xofy}
  \eeq
 in agreement with Eq. (4.2) of ref. \cite{BEACFARVOG}. The last equation is useful in obtaining the differential cross section (with respect to the recoil energy) after folding with the neutrino spectrum
\section {Coherent neutrino nucleus scattering}
\begin{figure}[!ht]
 \begin{center}
 \rotatebox{90}{\hspace{1.0cm} {$F^2(T_A) \rightarrow $}}
\includegraphics[scale=1.0]{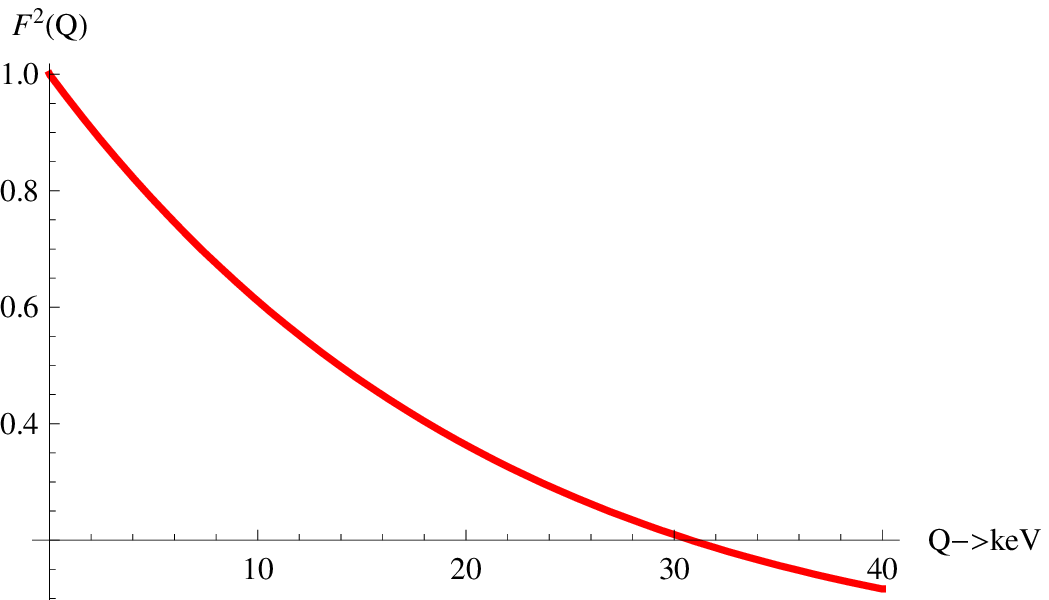}
\hspace{8.0cm}$T_A \rightarrow$ keV\\
 \rotatebox{90}{\hspace{1.0cm} {$F^2(T_A) \rightarrow $}}
\includegraphics[scale=1.0]{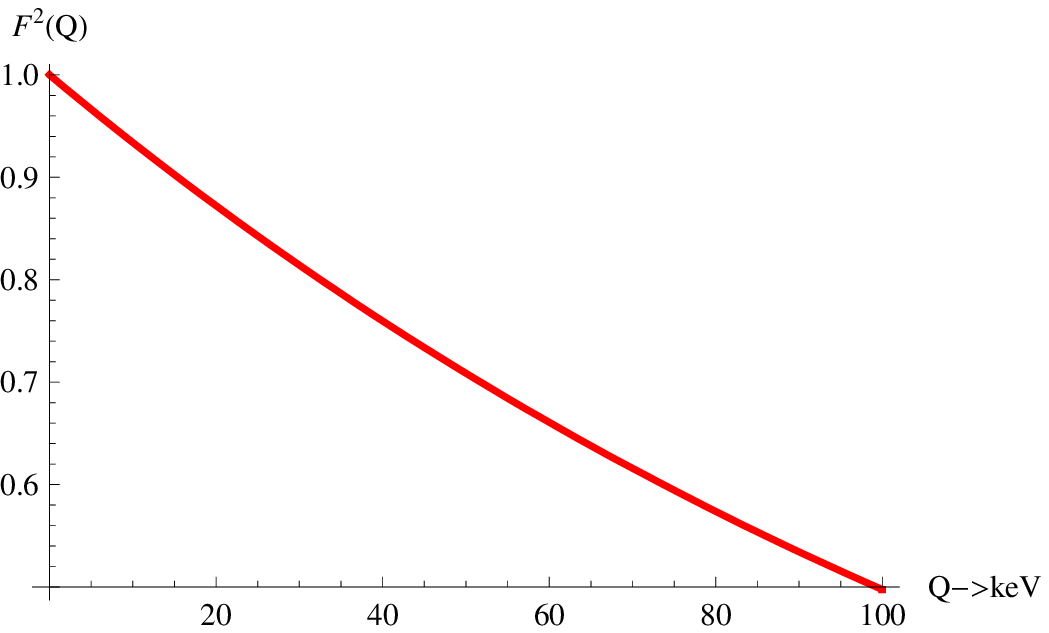}
\hspace{8.0cm}$T_A \rightarrow$ keV
 \caption{The square of the nuclear form factor, $F^2(T_A)$, as a function of the recoil energy for A=131 (top)
and A=40 (bottom). Note that the maximum recoil energy is different for each target.}
 \label{fig:ff}
 \end{center}
  \end{figure}

From the above expressions we see that the vector current contribution, which may lead to coherence, is negligible
in the case of the protons. Thus the coherent contribution \cite{PASCHOS} may come from the neutrons and is expected to be
proportional to the square of the neutron number.
The neutrino-nucleus scattering can be derived in analogous fashion. It can also be obtained from the amplitude of the neutrino nucleon scattering
%under the following assumptions:
%\begin{itemize}
%\item
by  employing the appropriate kinematics, i.e. those involving the elastically scattered nucleus and
%\item Ignore  effects of the nuclear form factor. Such effects, which are not expected to be very large,
% are currently under study and they will
 %appear elsewhere.
%\item The effective neutrino-nucleon amplitude is obtained in a similar fashion. It can also be obtained from the
 the substitution
$${\bf q}\Rightarrow \frac{{\bf p}}{A}~~,~~E_N \Rightarrow \sqrt{m_N^2+\frac{{\bf p}^2}{A^2}}=\frac{E_A}{A}$$
with ${\bf q}$ the nucleon momentum, $A$ is the nuclear mass number and ${\bf p}$ the nuclear momentum.
%\end{itemize}
Under the above assumptions the neutrino-nucleus cross section takes the form:
 \begin{eqnarray}
 \left(\frac{d\sigma}{dT_A}\right)&=&\frac{G^2_F Am_N}{2 \pi}
 [(M_V+M_A)^2 \left (1+\frac{T_A}{E_{\nu}} \right )
 \nonumber\\
&+ &(M_V-M_A)^2
(1-\frac{T_A}{E_{\nu}})^2
%\nonumber\\
+ (M_A^2-M_V^2)\frac{Am_NT_A}{E^2_{\nu}} ]
 \label{elaswA}
  \end{eqnarray}
  where $M_V$ and $M_A$ are the nuclear matrix elements associated with the vector and the axial currents
  respectively and $T_A$ is the energy of the recoiling nucleus.
 The axial current contribution vanishes for $0^+ \Rightarrow 0^+$ transitions.
Anyway, it is negligible in the case of coherent scattering on neutrons. Thus Eq. (\ref{elaswA})  is reduced to:
%   \begin{eqnarray}
\beq
 \left(\frac{d\sigma}{dT_A}\right)_{weak}=\frac{G^2_F Am_N}{2 \pi}~(N^2/4) F_{coh}(T_A,E_{\nu}),
%\nonumber\\
 \label{elaswAV1}
\eeq
with
\beq
F_{coh}(T_A,E_{\nu})= F^2(q^2)
  \left ( 1+(1-\frac{T_A}{E_{\nu}})^2
-\frac{Am_NT_A}{E^2_{\nu}} \right)
 \label{elaswAV2}
  \eeq
  where $N$ is the neutron number and $F(q^2)= F(T_A^2+2 A m_N T_A)$ is the nuclear form factor.
  The effect of the nuclear form factor depends on the target, since the maximum recoil energy depends
  on the target (see Fig. \ref{fig:ff}).
% and $Q_{fac}(T_A)$ the quenching factor.
  \section{Quenching factors and Energy thresholds }.
The above results refer to an ideal detector operating down to zero energy threshold. For a real detector, however,
as we have already mentioned, the nuclear recoil events are quenched, especially at low energies.
The quenching
factor for a given detector  is the ratio of the signal height for a recoil track divided by that of an electron signal with the same energy. We should not forget that the signal heights depend on the
velocity and how the signals are extracted experimentally. The actual quenching
  factors must be determined experimentally for each target. In the case of NaI the quenching
factor is 0.05, while for Ge and Si it is 0.2-0.3. For our purposes it is adequate, to multiply
%Eq. \ref{elaswAV2}
the energy scale by an recoil energy dependent quenching factor, $ Q_{fac}(T_A)$
  adequately described by the Lidhard theory \cite{LIDHART}.  More specifically in our estimate of $Qu(T_A)$ we assumed a quenching factor of the following empirical form \cite{LIDHART}, \cite{SIMON03}:
\beq
Q_{fac}(T_A)=r_1\left[ \frac{T_A}{1keV}\right]^{r_2},~~r_1\simeq 0.256~~,~~r_2\simeq 0.153
\label{quench1}
\eeq
\begin{figure}[!ht]
 \begin{center}
 \rotatebox{90}{\hspace{1.0cm} {$f_{\mbox{quench}}(T_A)\rightarrow $}}
\includegraphics[scale=0.8]{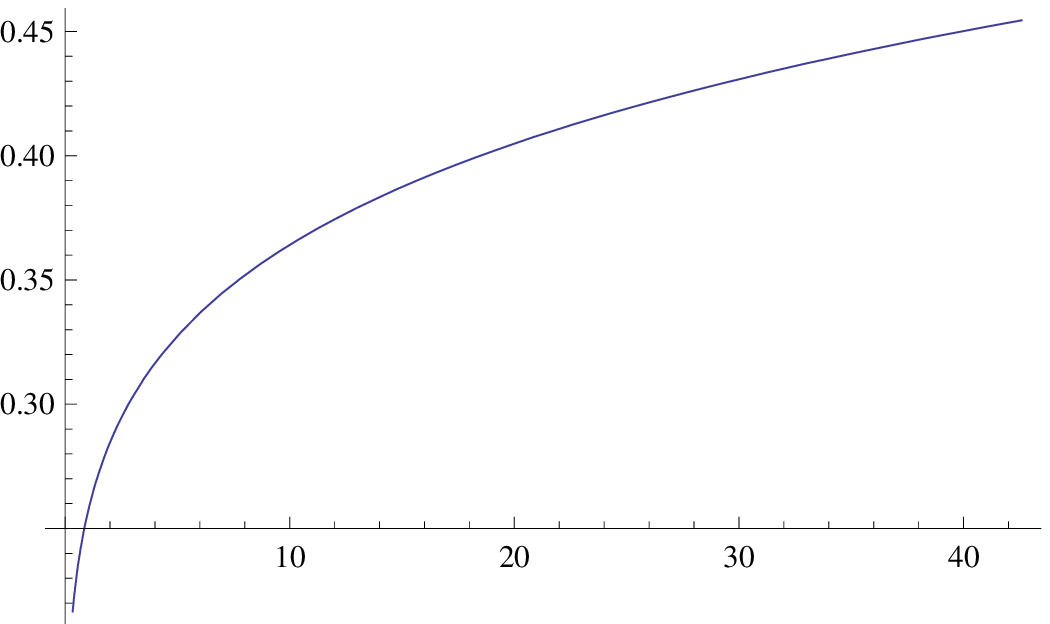}
\hspace{8.0cm}$T_A\rightarrow$ keV\\
 \rotatebox{90}{\hspace{1.0cm} {$f_{\mbox{quench}}(T_A)\rightarrow $}}
\includegraphics[scale=0.8]{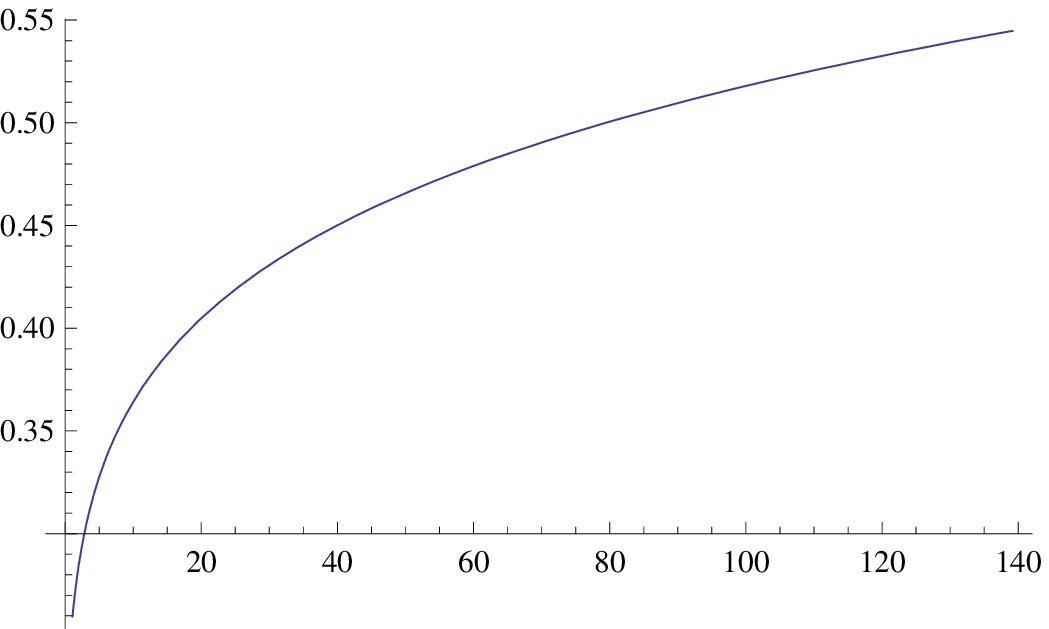}
\hspace{8.0cm}$T_A\rightarrow$ keV
 \caption{The quenching factor as a function of the recoil energy in the case of A=131 (top) and A=40 (bottom).}
 \label{fig:quench}
 \end{center}
  \end{figure}
  The quenching factors
very much depend on the detector type.
The quenching factor,exhibited in Fig. \ref{fig:quench} for recoil energies of $^{131}$Xe
and $^{40}$Ar, were obtained assuming the same quenching of the form of Eq. (\ref{quench1}).
In the presence of the quenching factor as given by Eq.( \ref{quench1})
the measured recoil energy is typically reduced by factors of about 3, when compared with the electron energy. In other words
a threshold energy of electrons of 1 keV becomes 3 keV for nuclear recoils.
\section{results}
\subsection{For a heavy target, $^{131}$Xe}
\begin{figure}[!ht]
 \begin{center}
  \rotatebox{90}{\hspace{1.0cm} {$\frac{d\sigma}{dT_A} \rightarrow 10^{-41}$ cm$^2$~keV$^{-1}$}}
% \rotatebox{90}{\hspace{1.0cm} {$f_{\mbox{quench}}(T_A)\rightarrow $}}
\includegraphics[scale=1.0]{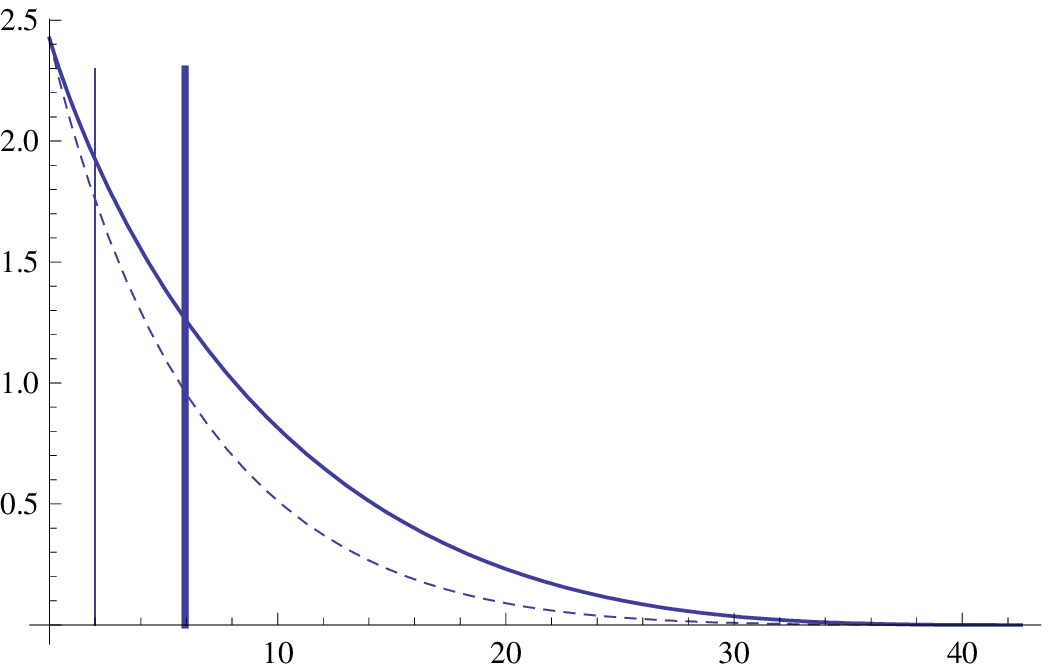}
\hspace{8.0cm}$T_A\rightarrow$ keV\\
 \rotatebox{90}{\hspace{1.0cm} {$\frac{d\sigma}{dT_A} \rightarrow 10^{-41}$ cm$^2$~keV$^{-1}$}}
% \rotatebox{90}{\hspace{1.0cm} {$f_{\mbox{quench}}(T_A)\rightarrow $}}
\includegraphics[scale=1.0]{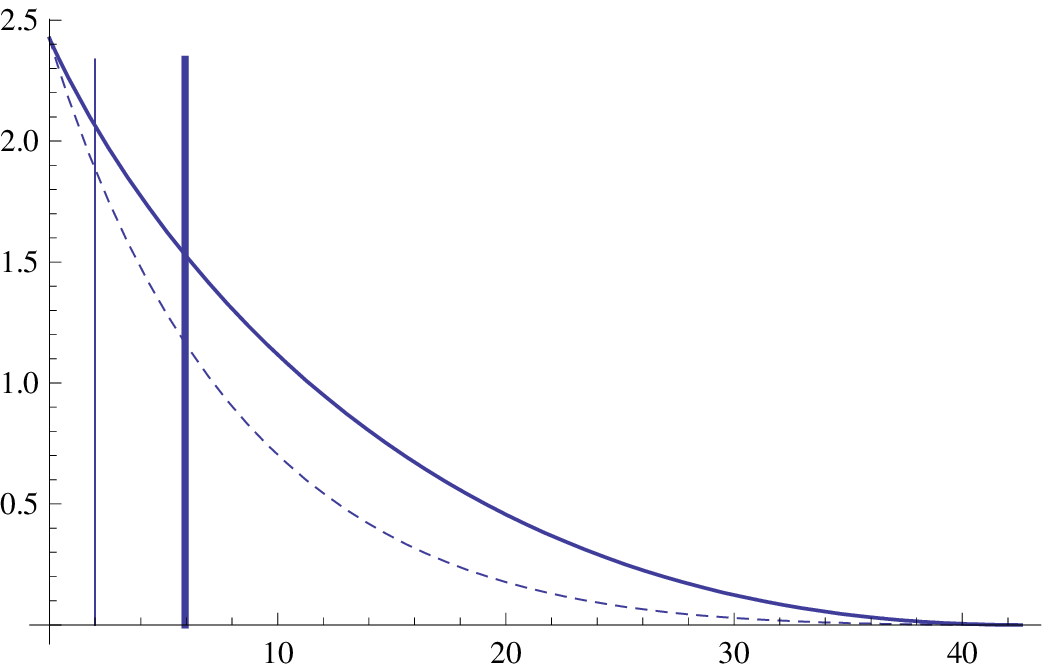}
  \hspace{8.0cm}$T_A\rightarrow$ keV\\
   \caption{The differetial neutrino nucleus cross section as a function of the recoil energy in the case of A=131 for the continuous $\nu_e$ spectrum  (top) and the continuous $\bar{\nu}_{\mu}$ spectrum  (bottom). The solid (dashed) curves correspond to no form factor (form factor) respectively. The fine vertical line corresponds to a threshold of 3 keV. The phase space on its left
is unavailable. Due to the presence of quenching the  excluded phase space becomes larger (on the left of the thick line).}
 \label{fig:dsigmadT131}
 \end{center}
  \end{figure}

\begin{figure}[!ht]
 \begin{center}
 \rotatebox{90}{\hspace{1.0cm} {$\frac{d\sigma}{dT_A} \rightarrow 10^{-41}$ cm$^2$~keV$^{-1}$}}
% \rotatebox{90}{\hspace{1.0cm} {$f_{\mbox{quench}}(T_A)\rightarrow $}}
\includegraphics[scale=1.0]{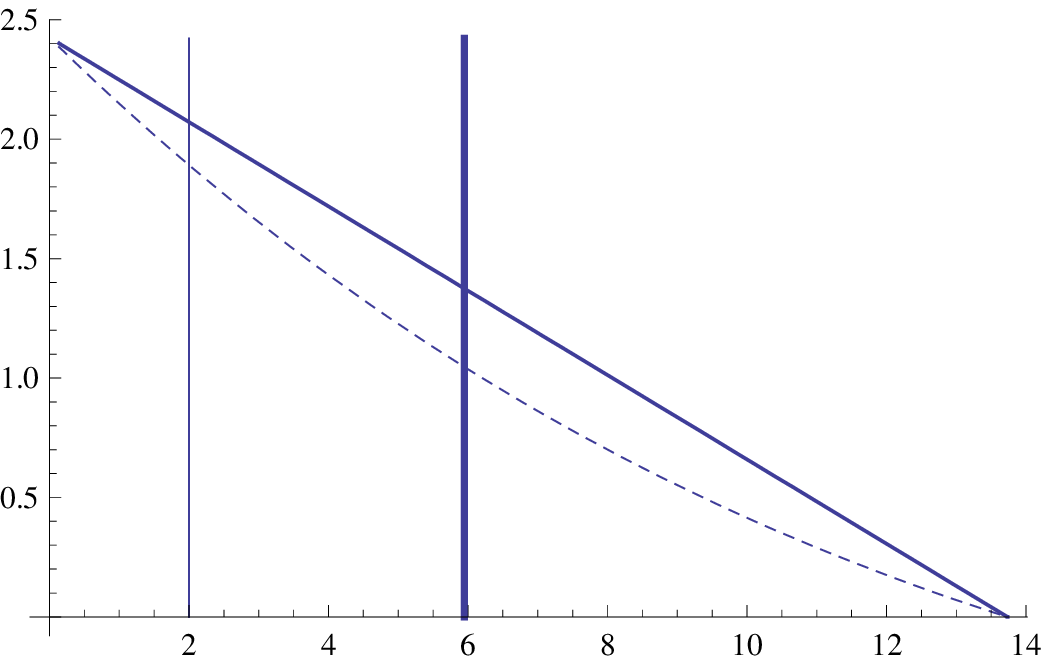}
\hspace{8.0cm}$T_A\rightarrow$ keV
 \caption{The same as in Fig.  \ref{fig:dsigmadT131} for the discreet $\nu_{\mu}$ spectrum.}
 \label{fig:dsigmadTd131}
 \end{center}
  \end{figure}

\begin{figure}[!ht]
 \begin{center}
%  \rotatebox{90}{\hspace{1.0cm} {$\frac{d\sigma}{dT_A} \rightarrow 10^{-42}$ cm$^2$~keV$^{-1}$}}
% \rotatebox{90}{\hspace{1.0cm} {$f_{\mbox{quench}}(T_A)\rightarrow $}}
\includegraphics[scale=1.0]{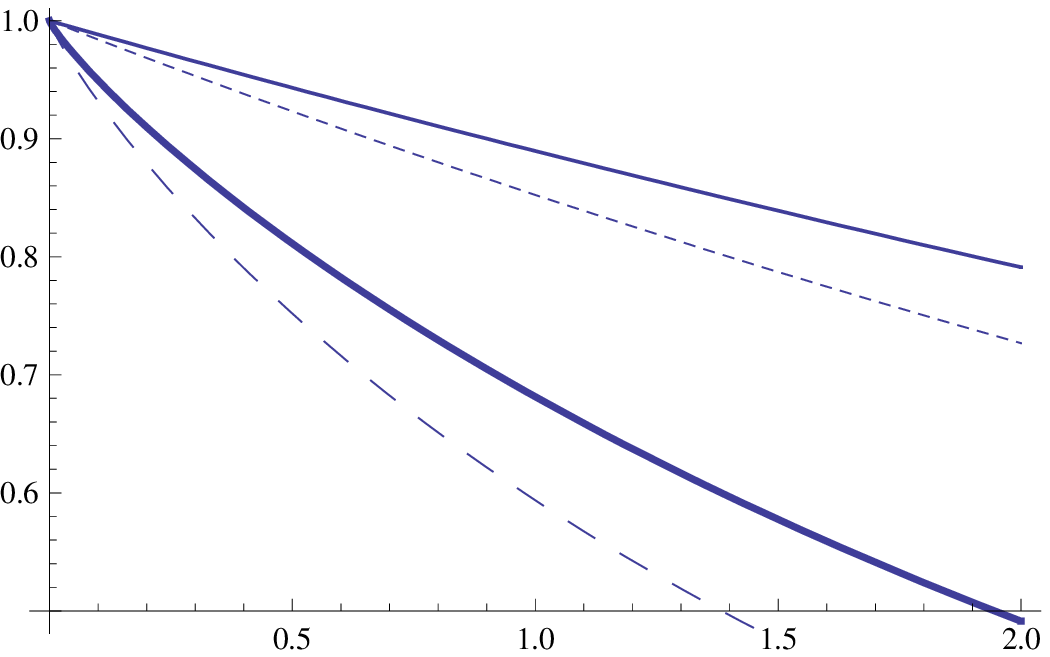}
\hspace{8.0cm}$E_{th}\rightarrow$ keV\\
% \rotatebox{90}{\hspace{1.0cm} {$\frac{d\sigma}{dT_A} \rightarrow 10^{-42}$ cm$^2$~keV$^{-1}$}}
% \rotatebox{90}{\hspace{1.0cm} {$f_{\mbox{quench}}(T_A)\rightarrow $}}
\includegraphics[scale=1.0]{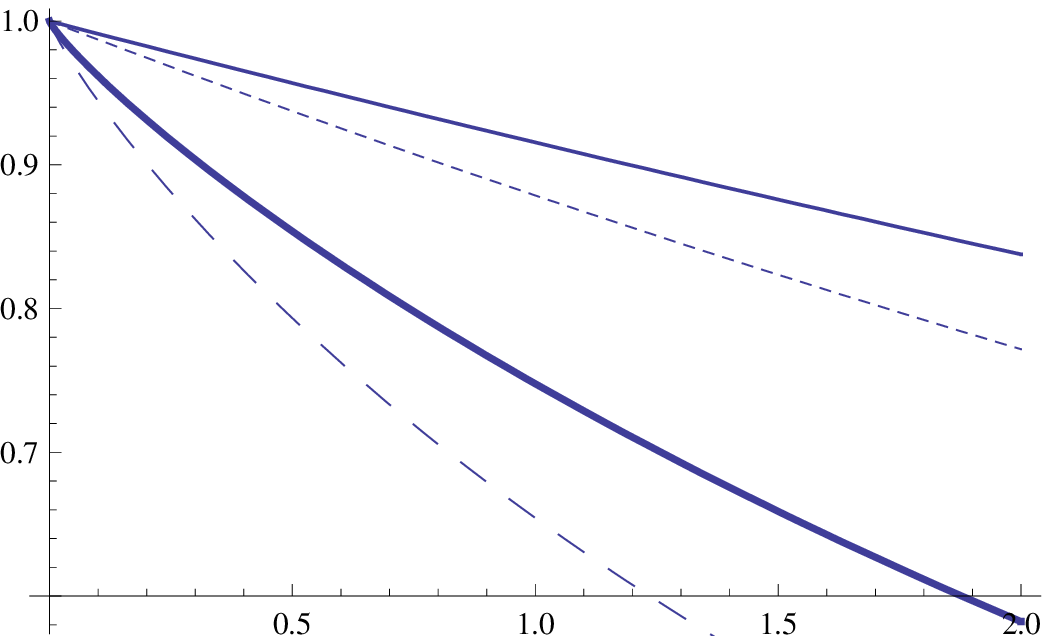}
\hspace{8.0cm}$E_{th}\rightarrow$ keV\\
 \caption{The event rates at a given energy threshold divided by the corresponding ones at zero threshold as a function of the thrshold energy in the case of A=131 target for the continuous $\nu_e$ spectrum  (top) and the continuous $\bar{\nu}_{\mu}$ spectrum  (bottom). The solid (dashed) curves correspond to no form factor
(form factor) respectively. The two lowest curves take into account to the quenching factor.}
 \label{fig:sigmath131}
 \end{center}
  \end{figure}

\begin{figure}[!ht]
 \begin{center}
 \rotatebox{90}{\hspace{1.0cm} {$\frac{d\sigma}{dT_A} \rightarrow 10^{-42}$ cm$^2$~keV$^{-1}$}}
% \rotatebox{90}{\hspace{1.0cm} {$f_{\mbox{quench}}(T_A)\rightarrow $}}
\includegraphics[scale=1.0]{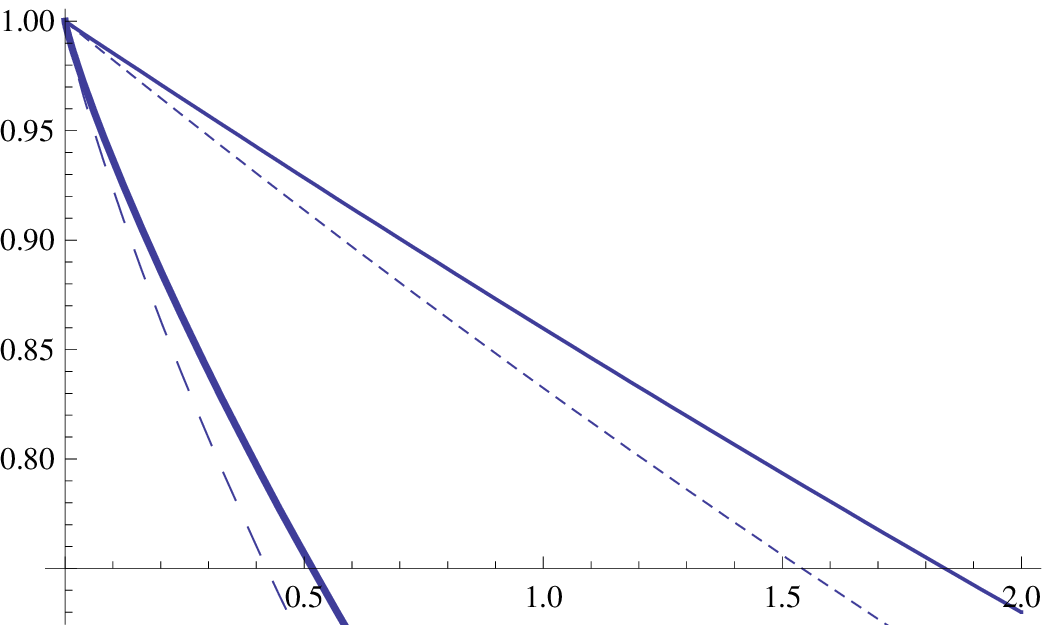}
\hspace{8.0cm}$E_{th}\rightarrow$ keV
 \caption{The same as in Fig.  \ref{fig:sigmath131} for the discreet $\nu_{\mu}$ spectrum.}
 \label{fig:sigmathd131}
 \end{center}
  \end{figure}

The neutrino-nucleus cross sections are shown in Figs \ref{fig:dsigmadT131} and \ref{fig:dsigmadTd131}. The quenching factor does not affect the rate (see Figs \ref{fig:dsigmadT131} and  \ref{fig:dsigmadTd131}) but it shifts the threshold downwards. In other words the unavailable phase
space is on the left of the thin line with no quenching and the thick line with quenching.  Clearly the effect
of quenching is going to be larger when  the recoil energy is smaller.
  Integrating the differential rates shown Figs \ref{fig:dsigmadT131} and \ref{fig:dsigmadTd131} at zerow threshold
  we obtain the total rates shown in table \ref{table:totalsigma}.
  \begin{table}[t]
\caption{The neutrino-nucleus total cross section in units of $10^{-41}$cm$^{-2}$due to the neutral current}
\label{table:totalsigma}
\begin{center}
\begin{tabular}{|c|c|c|c|c|c|c|}
\hline
\hline
 &   &  &  & & & \\
 target&$\nu_e$ &$\nu_e$  & $\bar{\nu}_{\mu}$& $\bar{\nu}_{\mu}$ &$\nu_{\mu}$ &$\nu_{\mu}$\\
  & (no FF)&  (FF)& (no FF)& (FF)& (no FF))& (FF)\\
  \hline
  $^{131}$Xe&2.71&1.98&3.60&2.46&2.79&1.79\\
  \hline
    $^{40}$Ar&0.221&0.190&0.294&0.244&0.178&0.162\\
   \hline
      \hline
%& $\times 10^{45}$cm$^{-2}$& $\times 10^{45}$cm$^{-2}$&$\times 10^{45}$cm$^{-2}$&$\times 10^{45}$cm$^{-2}$&$\times 10^{45}$cm$^{-2}$&$\times 10^{45}$cm$^{-2}$\\
%\end{tabular}
\end{tabular}
\end{center}
\end{table}
  The effect of a non zero threshold is exhibited in Figs  \ref{fig:sigmath131} and \ref{fig:sigmathd131}.
  It is clear that both the effect of the inclusion of the form factor and of quenching are important,
   especially in the case of high energy threshold,
  but the effect of quenching is most important of the two.
\subsection{For a light target, $^{40}$Ar}
\begin{figure}[!ht]
 \begin{center}
  \rotatebox{90}{\hspace{1.0cm} {$\frac{d\sigma}{dT_A} \rightarrow 10^{-42}$ cm$^2$~keV$^{-1}$}}
% \rotatebox{90}{\hspace{1.0cm} {$f_{\mbox{quench}}(T_A)\rightarrow $}}
\includegraphics[scale=1.0]{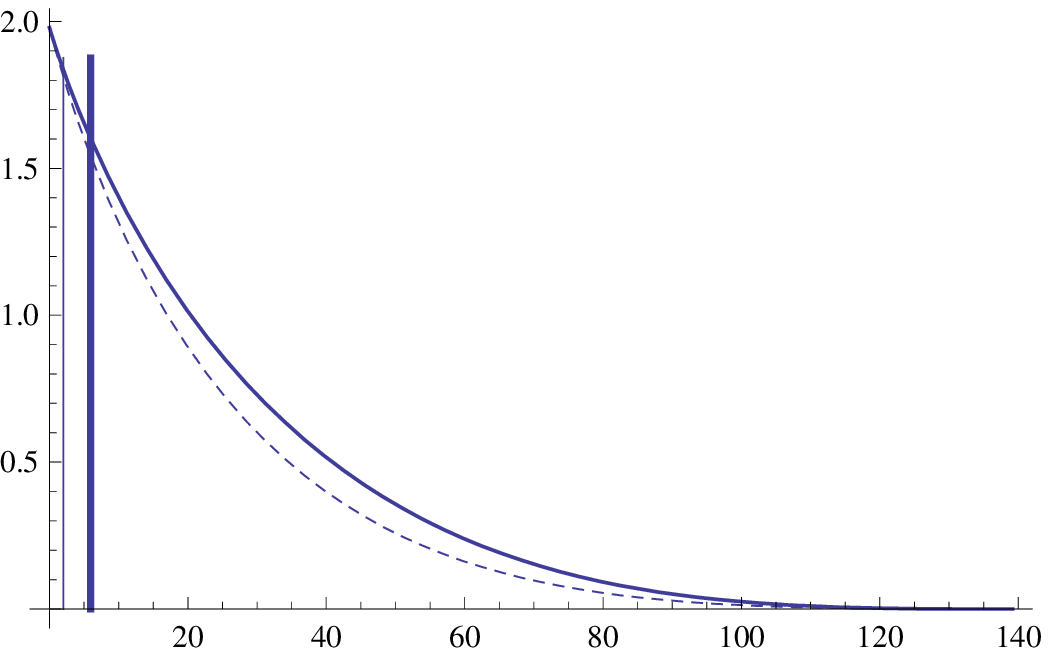}
\hspace{8.0cm}$T_A\rightarrow$ keV\\
 \rotatebox{90}{\hspace{1.0cm} {$\frac{d\sigma}{dT_A} \rightarrow 10^{-42}$ cm$^2$~keV$^{-1}$}}
% \rotatebox{90}{\hspace{1.0cm} {$f_{\mbox{quench}}(T_A)\rightarrow $}}
\includegraphics[scale=1.0]{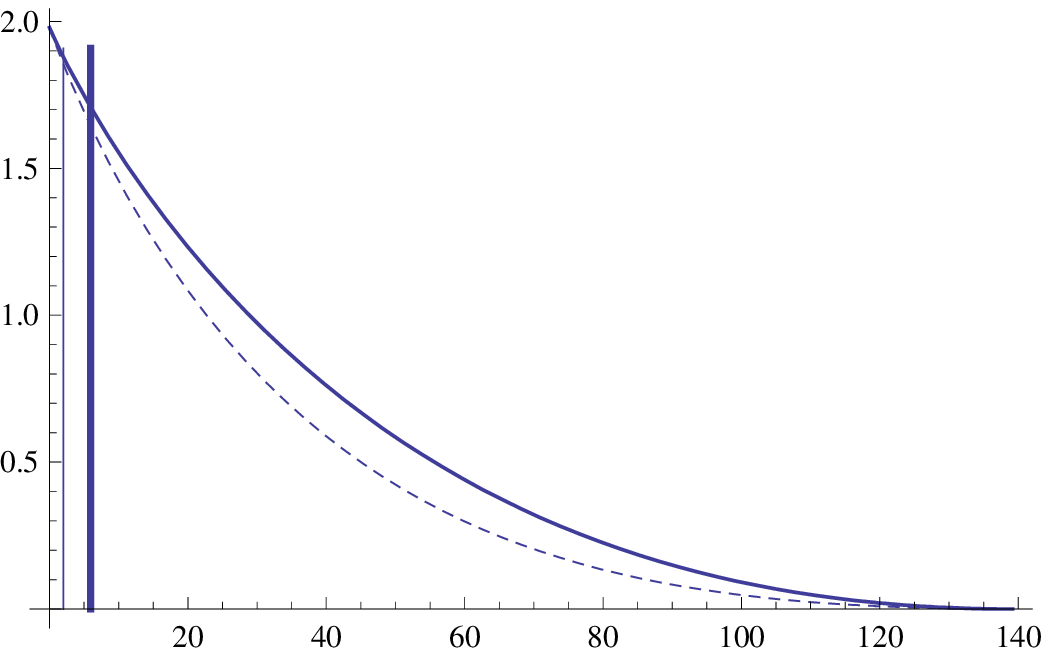}
 \hspace{8.0cm}$T_A\rightarrow$ keV\\
 \caption{The differential neutrino nucleus cross section as a function of the recoil energy in the case of A=40 for the continuous $\nu_e$ spectrum  (top) and the continuous $\bar{\nu}_{\mu}$ spectrum  (bottom). The solid (dashed) curves correspond to no form factor (form factor) respectively. The fine vertical line corresponds to a threshold of 3 keV. The phase space on its left
is unavailable. Due to the presence of quenching the  excluded phase space becomes larger (on the left of the thick line).}
 \label{fig:dsigmadT40}
 \end{center}
  \end{figure}

\begin{figure}[!ht]
 \begin{center}
 \rotatebox{90}{\hspace{1.0cm} {$\frac{d\sigma}{dT_A} \rightarrow 10^{-42}$ cm$^2$~keV$^{-1}$}}
% \rotatebox{90}{\hspace{1.0cm} {$f_{\mbox{quench}}(T_A)\rightarrow $}}
\includegraphics[scale=1.0]{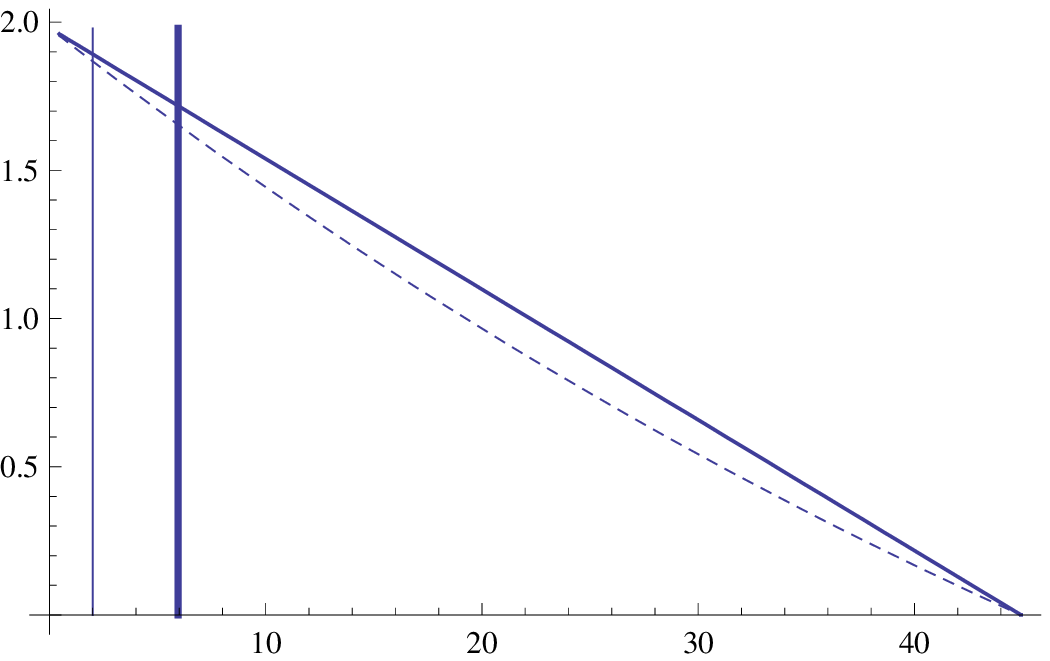}
\hspace{8.0cm}$T_A\rightarrow$ keV
 \caption{The same as in Fig.  \ref{fig:dsigmadT40} for the discreet $\nu_{\mu}$ spectrum.}
 \label{fig:dsigmadTd40}
 \end{center}
  \end{figure}

\begin{figure}[!ht]
 \begin{center}
%  \rotatebox{90}{\hspace{1.0cm} {$\frac{d\sigma}{dT_A} \rightarrow 10^{-42}$ cm$^2$~keV$^{-1}$}}
% \rotatebox{90}{\hspace{1.0cm} {$f_{\mbox{quench}}(T_A)\rightarrow $}}
\includegraphics[scale=1.0]{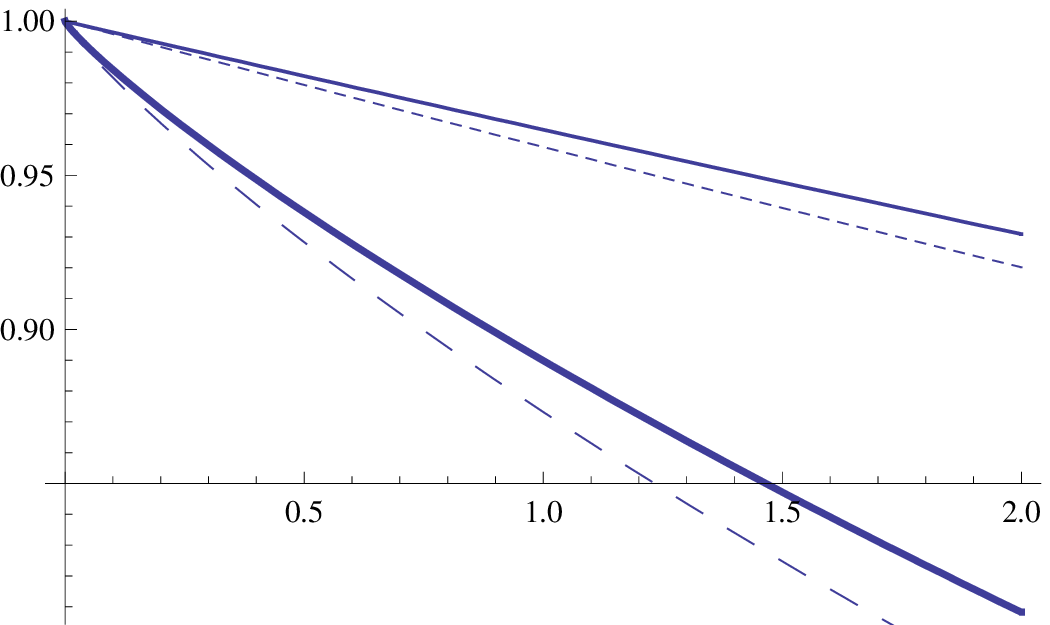}
\hspace{8.0cm}$E_{th}\rightarrow$ keV\\
% \rotatebox{90}{\hspace{1.0cm} {$\frac{d\sigma}{dT_A} \rightarrow 10^{-42}$ cm$^2$~keV$^{-1}$}}
% \rotatebox{90}{\hspace{1.0cm} {$f_{\mbox{quench}}(T_A)\rightarrow $}}
\includegraphics[scale=1.0]{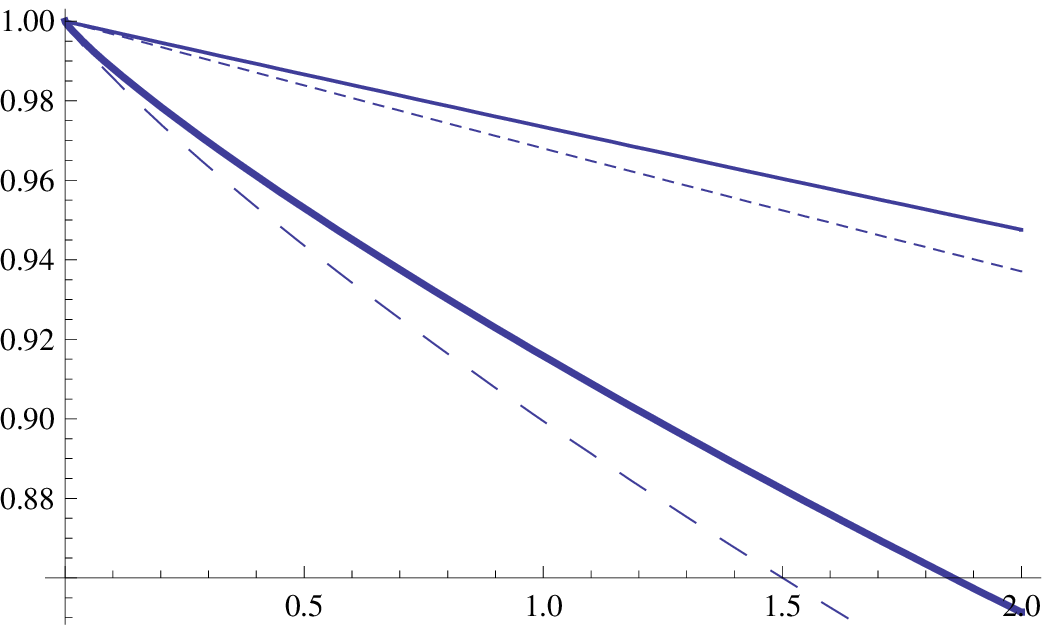}
\hspace{8.0cm}$E_{th}\rightarrow$ keV\\
 \caption{The event rates at a given energy threshold divided by the corresponding ones at zero threshold as a function of the thrshold energy in the case of A=40 target for the continuous $\nu_e$ spectrum  (top) and the continuous $\bar{\nu}_{\mu}$  spectrum  (bottom). The continuous (dashed) curves correspond to no form
factor (form factor) respectively. In each figure the thick solid and the long dash  curves  also take into account the quenching factor.}
 \label{fig:sigmath40}
 \end{center}
  \end{figure}

\begin{figure}[!ht]
 \begin{center}
% \rotatebox{90}{\hspace{1.0cm} {$\frac{d\sigma}{dT_A} \rightarrow 10^{-42}$ cm$^2$~keV$^{-1}$}}
% \rotatebox{90}{\hspace{1.0cm} {$f_{\mbox{quench}}(T_A)\rightarrow $}}
\includegraphics[scale=1.0]{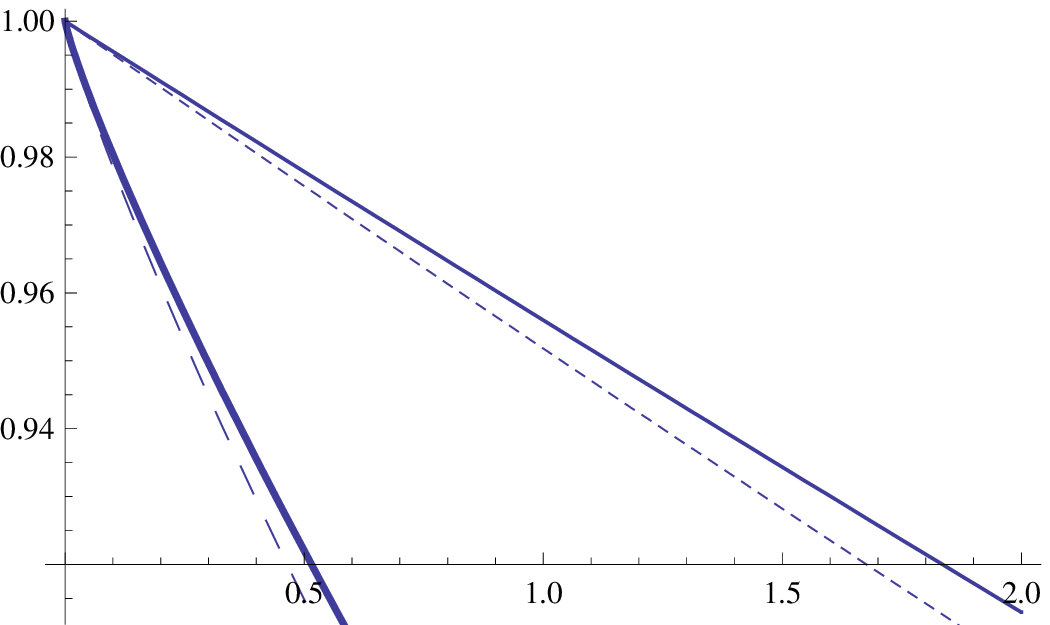}
\hspace{8.0cm}$E_{th}\rightarrow$ keV
 \caption{The same as in Fig.  \ref{fig:sigmath40} for the discreet $\nu_{\mu}$  spectrum.}
 \label{fig:sigmathd40}
 \end{center}
  \end{figure}

The differential cross sections are now shown in Figs \ref{fig:dsigmadT40} and \ref{fig:dsigmadTd40}.
  Integrating the differential rates shown Figs \ref{fig:dsigmadT40} and \ref{fig:dsigmadTd40} at zero threshold
  we obtain the total rates shown in table \ref{table:totalrates}.
  The effect of a non zero threshold is exhibited in Figs  \ref{fig:sigmath40} and \ref{fig:sigmathd40}.
  The effect of quenching now is less dramatic since the recoil energy for a light target is higher.
  The total number of events for a  target of mass $m$ in time $t$ is given by
  \beq
  R=3.156\times 10^7\frac{m_t}{1 kg}\frac{t}{1y}\Phi(\nu,L)\sigma(\nu,(A,N)) \frac{1000}{A_t} N_A
  \eeq
  where $N_A=6.02\times 10^{23}$ is Avogadro's number, $A_t$ the mass number of the target,
  $ \sigma(\nu,(A,N))$ is the coherent neutrino-nucleus cross section and
$ \Phi(\nu,L)$ is the neutrino flux (neutrinos per second  per cm$^2$)
  in distance L from the source. Using
$$ \Phi(\nu,L=50\mbox{m})=1.96\times 10^{6}\mbox{cm}^{-2}\mbox{s}^{-1}$$
for each flavor we obtain
  the data of table \ref{table:totalrates}.
 \section{Experimental issues with the spherical TPC}
\begin{figure}[!ht]
 \begin{center}
%  \rotatebox{90}{\hspace{1.0cm} {$\frac{d\sigma}{dT_A} \rightarrow 10^{-42}$ cm$^2$~keV$^{-1}$}}
 \rotatebox{90}{\hspace{1.0cm} {$s(R,R/L)/s(10,0.2) \rightarrow $}}
\includegraphics[scale=1.0]{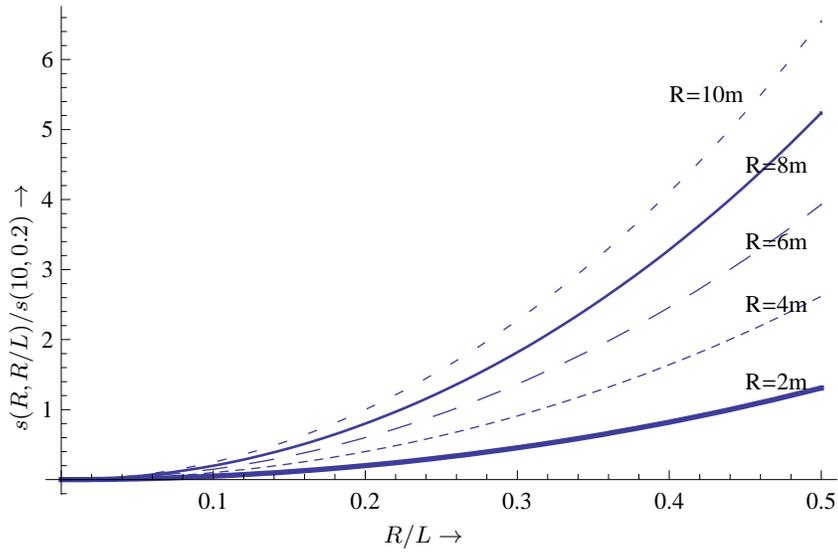}
\hspace{8.0cm}$R/L\rightarrow$ \\
 \caption{The parameter $s(R,R/L)$ (see text) in the case of a sphere of radius $R$ whose center is at a distance $L$ from the source.}
 \label{Fig:sphere}
 \end{center}
  \end{figure}

\begin{figure}[!ht]
 \begin{center}
%  \rotatebox{90}{\hspace{1.0cm} {$\frac{d\sigma}{dT_A} \rightarrow 10^{-42}$ cm$^2$~keV$^{-1}$}}
% \rotatebox{90}{\hspace{1.0cm} {$s(R,R/L)/s(10,0.2) \rightarrow $}}
\includegraphics[scale=0.9]{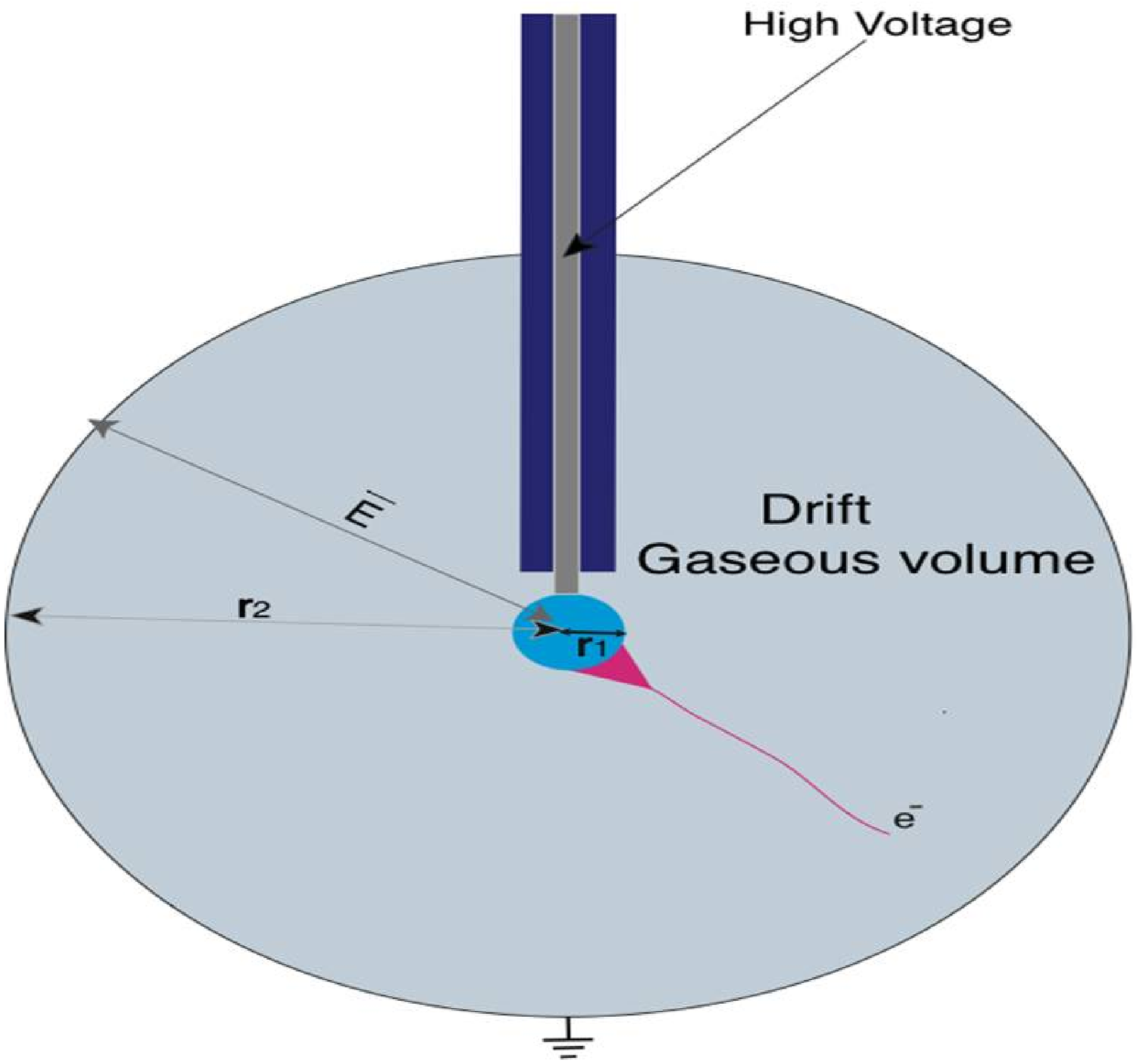}
%\hspace{8.0cm}$R/L\rightarrow$ \\
 \caption{A sketch of spherical TPC detector to be employed in studying the coherent scattering of neutrionos eployed in this work.}
 \label{Fig:sphere_app}
 \end{center}
  \end{figure}

\begin{figure}[!ht]
 \begin{center}
%  \rotatebox{90}{\hspace{1.0cm} {$\frac{d\sigma}{dT_A} \rightarrow 10^{-42}$ cm$^2$~keV$^{-1}$}}
% \rotatebox{90}{\hspace{1.0cm} {$s(R,R/L)/s(10,0.2) \rightarrow $}}
\includegraphics[scale=0.6]{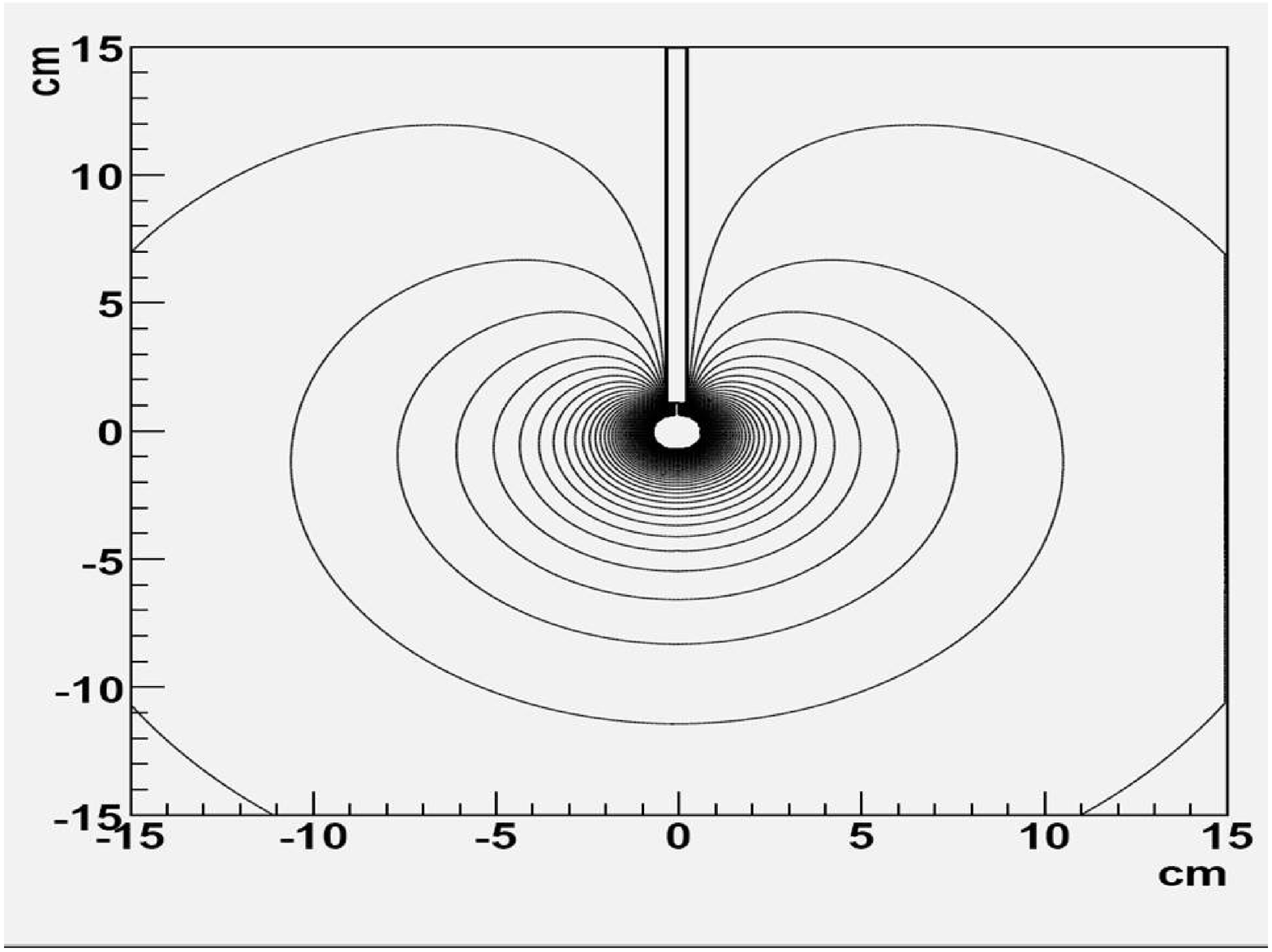}
\includegraphics[scale=0.6]{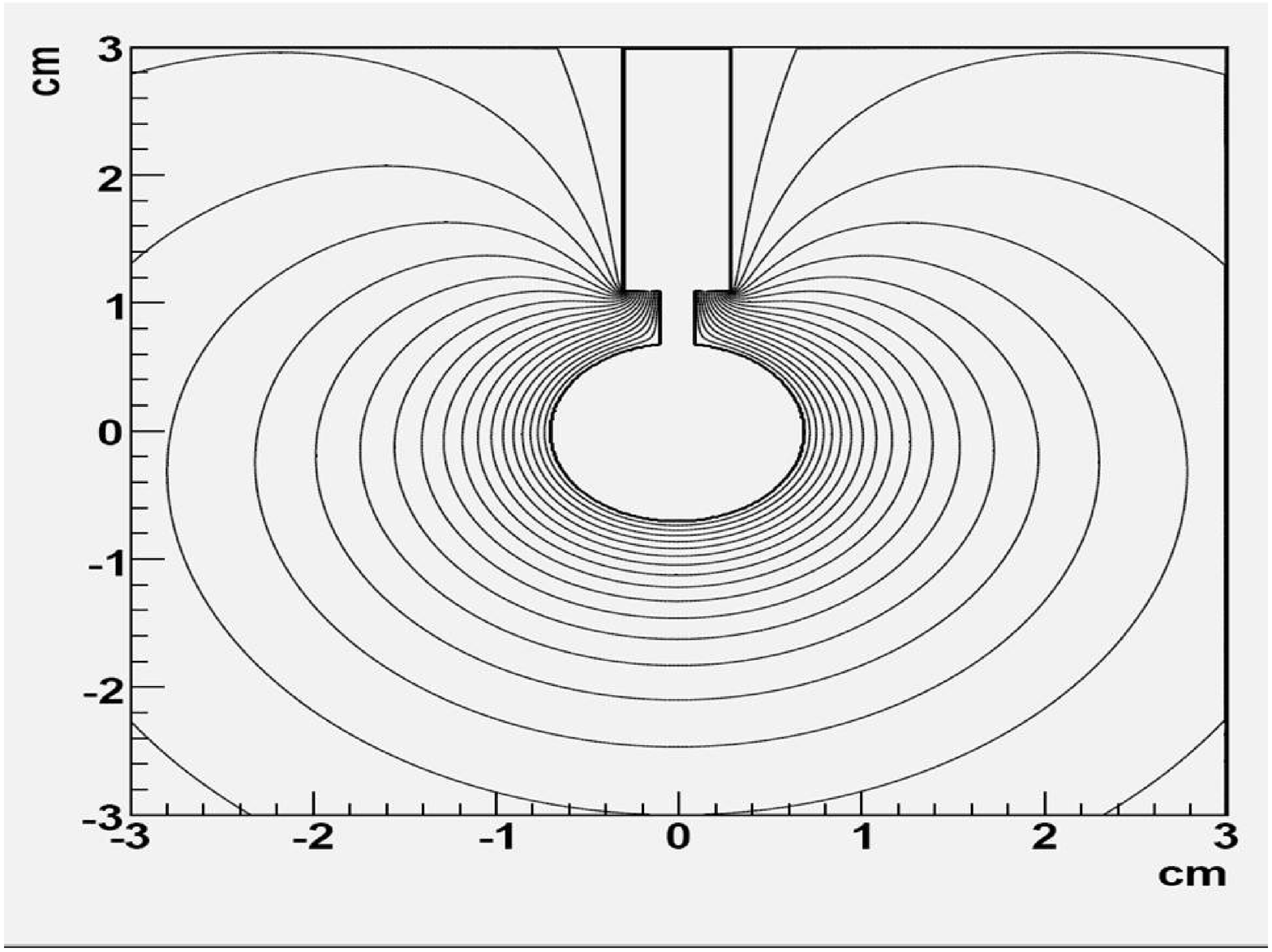}
%\hspace{8.0cm}$R/L\rightarrow$ \\
 \caption{ The Equipotential surfaces inside the spherical vessel of Fig.  \ref{Fig:sphere_app}. In spite of the wire connecting the inside and the outside spheres the field corrector introduced ensures that the electric field is spherically symmetric.}
 \label{Fig:sph1and2}
 \end{center}
  \end{figure}

The SNS facility provides an excellent opportunity to employ and test the spherical TPC gaseous detector \cite{sphere08} of volume $V$ under pressure $P$ and temperature $T$ filled with a noble gas. This detector has good resolution and very low energy threshold. In this case  we obtain:
$$ R= 3.156\times 10^7 \frac{t}{1y}\Phi(\nu,L)\sigma(\nu,(A,N)) \frac{PV}{kT} s(L,x) $$
 where the parameter $s(L,x)$ depends on the shape of the vessel and the distance $L$ of its center  from the source.
 Using
$$ \Phi(\nu,L=50\mbox{m})=1.96\times 10^{6}\mbox{cm}^{-2}\mbox{s}^{-1}$$
for each flavor we obtain
  the data of table \ref{table:totalrates2}. The function s(L,x) for a sphere of radius R with its center at a distance $L$ from the source is given by
\beq
s(R,R/L)=\frac{L^2}{(4 /3)\pi R^3} 2 \pi \int_{L-R}^{L+R} r^2 dr \int_ 0^{\pi } d \theta \sin{\theta}\frac{1}{R^2+r^2+2 r L \cos{\theta}}
\eeq
The ratio $s(R,R/L)/s(10,0.2$ is shown   and Fig.
\ref{Fig:sphere} .
    \begin{table}[t]
\caption{The event rates for nuclear recoils per ton per year due
to neutral current neutrino-nucleus scattering for a detector at a
distance L=50m away from the neutrino source with zero energy
threshold.
}
\label{table:totalrates}
\begin{center}
\begin{tabular}{|c|c|c|c|c|c|c|c|c|}
\hline
\hline
 &   &  &  & & & &&  \\
 target&$\nu_e$ &$\nu_e$  & $\bar{\nu}_{\mu}$& $\bar{\nu}_{\mu}$ &$\nu_{\mu}$ &$\nu_{\mu}$&all $\nu$&all $\nu$\\
  & (no FF)&  (FF)& (no FF)& (FF)& (no FF))& (FF)& (no FF))& (FF)\\
  \hline
  $^{131}$Xe&  7.70&5.64&10.3&6.99&6.21&5.09&24.0&17.69\\
  \hline
    $^{40}$Ar&2.06&1.77&2.74&2.27&1.66&1.51&6.46&5.5\\
   \hline
      \hline
%& $\times 10^{45}$cm$^{-2}$& $\times 10^{45}$cm$^{-2}$&$\times 10^{45}$cm$^{-2}$&$\times 10^{45}$cm$^{-2}$&$\times 10^{45}$cm$^{-2}$&$\times 10^{45}$cm$^{-2}$\\
%\end{tabular}
\end{tabular}
\end{center}
\end{table}
   \begin{table}[t]
\caption{The event rates for nuclear recoils  per year due
to neutral current neutrino-nucleus scattering for a detector with the shape of a sphere of radius $R=10$ m with its center at a distance $L=50$ m from the source(s(10,0.2)=1.008). The vessel is filled with gas under pressure $P=10$ atm and temperature $T=300~^0$K. The detector is assumed to operate at zero threshold.
}
\label{table:totalrates2}
\begin{center}
\begin{tabular}{|c|c|c|c|c|c|c|c|c|}
\hline
\hline
 &   &  &  & & & & &  \\
 target&$\nu_e$ &$\nu_e$  & $\bar{\nu}_{\mu}$& $\bar{\nu}_{\mu}$ &$\nu_{\mu}$ &$\nu_{\mu}$&all $\nu$&all $\nu$\\
  & (no FF)&  (FF)& (no FF)& (FF)& (no FF))& (FF)& (no FF))& (FF)\\
  \hline
  $^{131}$Xe&  1705&1249&2280&1548&1393&1120&5379&3917\\
  \hline
    $^{40}$Ar&139.3&119.7&185.3&153.5&112.2&102.1&436.86&375.3\\
   \hline
      \hline
\end{tabular}
\end{center}
\end{table}
The experimental set up is shown in Fig. \ref{Fig:sphere_app}.
Using an appropriate field corrector the field inside the detector is spherically symmetric (see Fig. \ref{Fig:sph1and2} ).
 \section{Discussion and Conclusions}
The total number of events per ton-y at 50 m is:
\beq
R(\mbox{Xe})=17.8\mbox{(events per ton-y)},~~~~R(\mbox{Ar})=5.5 \mbox{(events per ton-y)}
\eeq
One notices the following:
\begin{enumerate}
\item The content of  tables \ref{table:totalrates}-\ref{table:totalrates2} can be understood by comparing it to other neutrino sources.
 The flux at 50 m is $\Phi=3 \Phi_{\bar{\nu}_e}= 5.85\times 10^6 \mbox{cm}^{-2}\mbox{s}^{-1}$. This is comparable to  that of solar boron neutrinos   $\Phi_{\nu_e}= 5.0\times 10^6\mbox{cm}^{-2}\mbox{s}^{-1}$, but with a higher cross section, since the latter is characterized by lower average energy. Thus the relatively low flux and energy of the boron solar neutrinos  cannot cause any background problems to dark matter searches  \cite{VerEji08}. Anyway the number of neutrinos in a year is $N_{\nu}=1.8\times 10^{14}\mbox{cm}^{-2}$.  The corresponding quantities for supernovae neutrinos at a distance $D=3.1\times 10^{22}$ cm,  are  \cite{VERGIOM06}  $\Phi_{\nu_e}= 1.5\times 10^{12} \mbox{cm}^{-2}$,  $\Phi_{\bar{\nu}_e}= 1.0\times 10^{12} \mbox{cm}^{-2}$ and $\Phi_{\nu_{\alpha}}= 6.3 \times 10^{11} \mbox{cm}^{-2}$
for all other flavors $\alpha$, i.e a total of $3.1\times 10^{12} \mbox{cm}^{-2}$ . In other words, since the energy spectra are not very different, the event rate here is expected to be higher than that of a typical supernova in our galaxy\cite{VERGIOM06}.
\item The event rate per unit target mass  predicted does not favor the heavy target according to the expected $N^2$ rule implied by the differential cross section (compare Figs \ref{fig:dsigmadT131}-\ref{fig:dsigmadTd131} and \ref{fig:dsigmadT40}-\ref{fig:dsigmadTd40} ) for the following reasons:
\begin{itemize}
\item There fewer atoms in a kg of a heavy target. So one expects:
$$R \propto \frac{N^2}{A}.$$
\item The phase space available favors a light target, since, for a given neutrino energy, the nuclear recoil energy decreases with A. This kinematical advantage maybe partly offset by the fact that the effect of  form factor becomes  more important, if the recoil energy becomes higher.
\item Threshold effects may somewhat complicate the issue (compare Figs \ref{fig:sigmath131}-\ref{fig:sigmathd131} and
 \ref{fig:sigmath40}-\ref{fig:sigmathd40}).
\end{itemize}
 This $N^2$ dependence clearly applies  whenever the amount of the target material for a given volume is determined by the pressure (see table \ref{table:totalrates2}) in agreement  with our earlier  work on supernova neutrino detectors  \cite{VERGIOM06}.
\end{enumerate}
Acknowledgments:
One of the authors (JDV) is happy to acknowledge support by MRTN-CT-2006 (UniverseNet) and is  indebted to the CERN Theory Division for support and hospitality during the last phases of this work. One of us (FTA) was supported by the National Science Foundation Grant PHY-0500337.

%\bibliography{TeX}

\end{document}